\documentstyle[epsfig,longtable]{aipproc}
\def\3p0{$^3$P$_0$}
\begin{document}
\title{Photoproduction of Hybrid Mesons}

\author{T. Barnes}
\address{Physics Division, Oak Ridge National Laboratory \\
Oak Ridge, TN 37831-6373, and \\
Department of Physics , University of Tennessee \\
Knoxville, TN 37996-1501
}
\maketitle

\begin{abstract}
In this contribution I discuss prospects for photoproducing hybrid mesons
at CEBAF, based on recent model results and experimental indications
of possible hybrids. One excellent opportunity appears to be a search for $I=1$, 
$J^{PC}$ =
$2^{+-}$ 
``$b_2^o$''
hybrids in ($a_2\pi$)$^o$ through diffractive 
photoproduction. Other notable possibilities 
accessible through $\pi^+$ or $\pi^o$ exchange photoproduction are 
$I=1$ 
$1^{-+}$ 
``$\pi_1^+$''
in
$f_1\pi^+$,
$(b_1\pi)^+$
and
$(\rho\pi)^+$;
$\pi_J^+(1770)$
in
$f_2\pi^+$ and
$(b_1\pi)^+$;
$\pi^+(1800)$
in
$f_0\pi^+$, 
$f_2\pi^+$, 
$\rho^+\omega$ and
$(\rho\pi)^+$;
$a_1$ in 
$f_1\pi^+$ and
$f_2\pi^+$; and
$\omega$ in
$(\rho\pi)^o$, $\omega\eta$ and
$K_1K$.

\end{abstract}

\subsection*{Introduction: Expectations for hybrids}

The search for hybrids has reached an exciting and somewhat bewildering point.
Theoretical predictions for 
hybrid masses have historically been somewhat model dependent,
with masses for the lightest hybrid meson multiplet typically lying in the range
1.5-2.0 GeV. The flux-tube model prediction of 1.9 GeV \cite{hybr_ft_mass} 
has been the most widely
cited hybrid mass estimate. This appears to have been confirmed by recent
LGT studies \cite{hybr_latt} 
which find that the lightest exotic $n\bar n$-hybrid is a $1^{-+}$,
with a mass of about 2.0 GeV. Thus, theory appears to have reached
a consensus that exotic hybrids begin at around 2.0 GeV. The flux-tube model
predicts that the $I=1$ $1^{-+}$ should not be very broad; a width of about
$0.2$ GeV is anticipated for a mass near 1.9 GeV, with dominant decay
modes of $b_1\pi$ and $f_1\pi$ \cite{cp}.
One surprise from LGT is
that the 
$0^{+-}$ exotic is found by the MILC Collaboration to have a high mass,
perhaps about 2.7 GeV (from their figure). In the flux-tube model this should
be approximately degenerate with the $1^{-+}$.

These predictions are in striking {\it disagreement} with recent experimental
results~\cite{hybr_pi1}.
VES and E852 are in agreement about the phase motion of the $\eta\pi^-$
system in $\pi^- p \to \pi^- \eta n$,
which E852 notes can be fitted with a rather broad exotic
``$\pi_1(1400)$'' with 
$M=1.4$ GeV, 
$\Gamma_{tot}=0.4$ GeV. 
This mass is about $0.5$ GeV below theoretical expectations, and the
observed state is
{\it much} wider than the flux-tube model would anticipate for a
hybrid at this mass.
This state appears to have been confirmed by Crystal Barrel \cite{CBetapi}.
A second
$I=1$ $1^{-+}$ exotic has been found by E852 near 1.6 GeV in $\rho\pi$, 
with a width slightly below 0.2 GeV. This relatively narrow state
shows clear resonant phase motion against the 
$\pi_2(1670)$.

Thus hybrids may well have been discovered, and the disagreement with 
theoretical predictions, 
including LGT, 
makes their further study of the greatest interest. In this contribution
we 
consider which hybrids might be 
photoproduced easily (diffractively or by pion exchange) in a future
search for exotic mesons at CEBAF.

\subsection*{Channels for hybrid photoproduction}

Here we simply list accessible quantum numbers, since detailed
model calculations at this early stage may be inappropriate. The three
mechanisms believed to be the most important 
in light meson photoproduction \cite{tbphot} are diffractive
vector-nucleon scattering (vector dominance followed by pomeron exchange), 
$t$-channel meson exchange (with the
pion giving the leading contribution at small $t$), and baryon resonance decay.
The latter can be treated as a background, and the former two can be selectively
enhanced
through $t$ cuts and the selection of final state quantum numbers.

Diffractive scattering is poorly understood at the QCD level, although some
features are generally agreed on; it corresponds to vacuum quantum number
exchange, an imaginary amplitude, and may be dominated by $gg$ exchange.
A simple $I=0$, $0^{++}$ exchange picture gives the list of 
quantum numbers in Fig. 1. The accessible exotics are all $J^{PC} = even^{+-}$;
$I=1$, $G=(+)$ is favored due to the larger contribution of $\rho^o$ to vector
dominance.
\begin{figure}[b!] 
\centerline{\epsfig{file=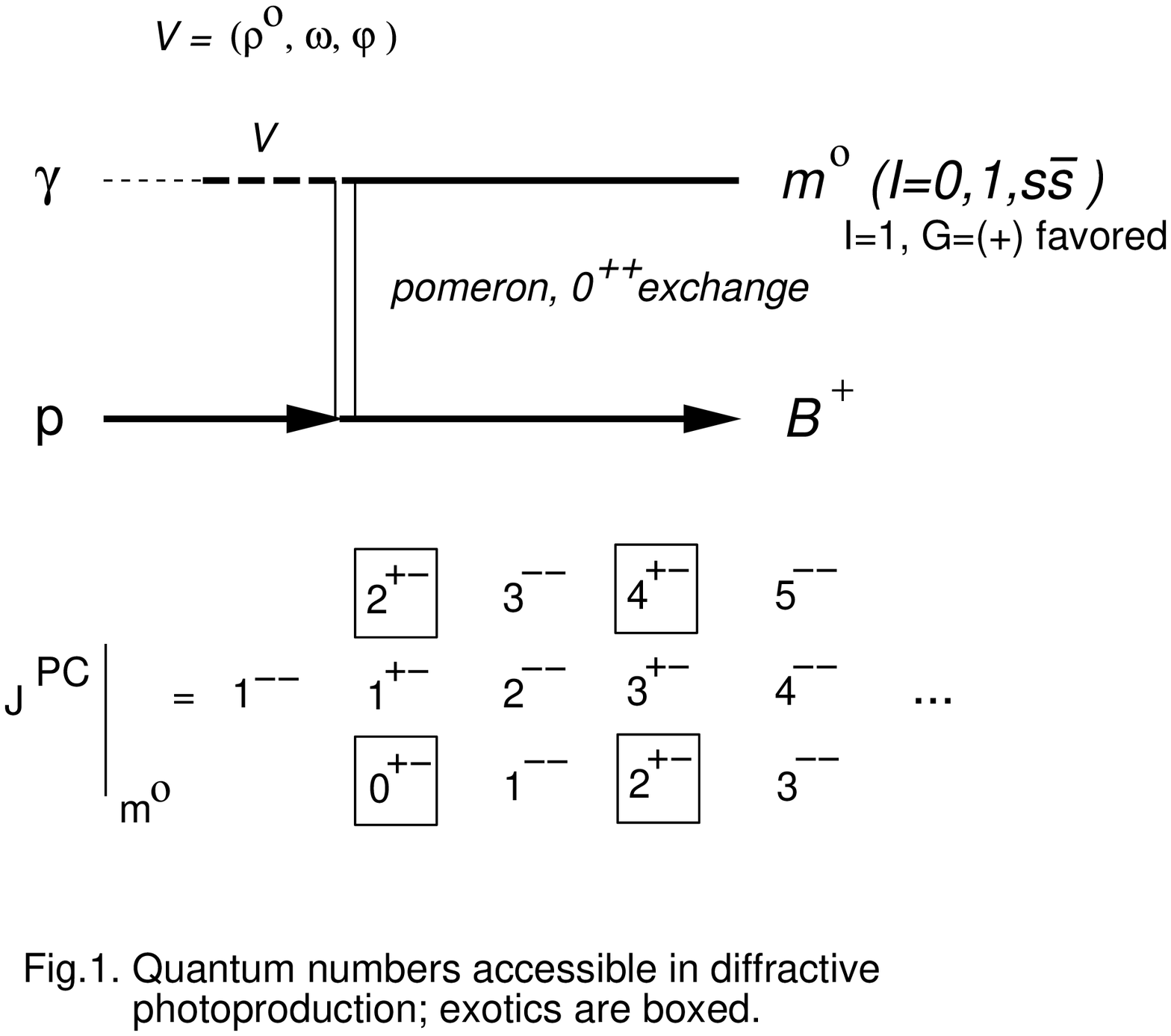,height=3in,width=3in}}
\vspace{10pt}
\end{figure}

Assuming that the flux-tube model \cite{hybr_ft_mass,cp} 
is a useful guide to masses and decays of hybrids,
the 
$0^{+-}$
and 
$2^{+-}$ exotics $b_0$ and $b_2$ are the most interesting because they
are in the lowest flux-tube hybrid multiplet.
The $b_0$ however is predicted to be extremely
broad. The $b_2$ is much narrower, with a predicted width of 0.4 GeV and
a 50\% branch to
$a_2\pi$. Diffractive photoproduction of the neutral $(a_2\pi)^o$ system 
therefore
appears to offer an excellent opportunity for identifying a hybrid in 
photoproduction. 

The $s\bar s$ $2^{+-}$ exotic can also be diffractively photoproduced,
and should decay mainly to $K^*_2K$ and $K_1K$ (both $K_1$ states).
In the flux-tube model this state has a width of 0.2 GeV and may 
be clearly evident because of smaller backgrounds in strange channels.

Pion exchange photoproduction offers several possibilities.
There is a {\it caveat} that S+S couplings of hybrids
in the flux-tube model are suppressed, so photoproducing a hybrid
through $\rho^0 + \pi$ might have a relatively weak amplitude. 
Of course the exotic candidates $\pi_1(1400)$ and 
$\pi_1(1600)$ show no S+S suppression, and 
violation of this selection rule in improved flux-tube calculations
is found to be
significant \cite{cp}.
One-pion exchange photoproduction of hybrids 
certainly merits investigation, independent of these 
flux-tube predictions
of possibly suppressed couplings.

First, for charged $\pi$ exchange we find the list of quantum numbers
shown in Fig.2. Here $I=1$ is forced, and $G=(-)$ is preferred. This gives
$odd^{-+}$ exotics, the most interesting of course being the 
$I=1$ $1^{-+}$. 
Here one would first study $\eta\pi$, $\eta'\pi$ and $\rho\pi$,
to see if the E852 states
$\pi_1(1400)$ and
$\pi_1(1600)$ are evident. Since the $\pi_1(1600)$ is reported to couple
strongly to $\rho\pi$, it should certainly appear in this reaction!
Earlier work on photoproduction by
Condo {\it et al.} \cite{Condo} found another possible
exotic, a $\pi_J(1770)$ state
which may be $1^{-+}$ or $2^{-+}$.
\begin{figure}[b!] 
\centerline{\epsfig{file=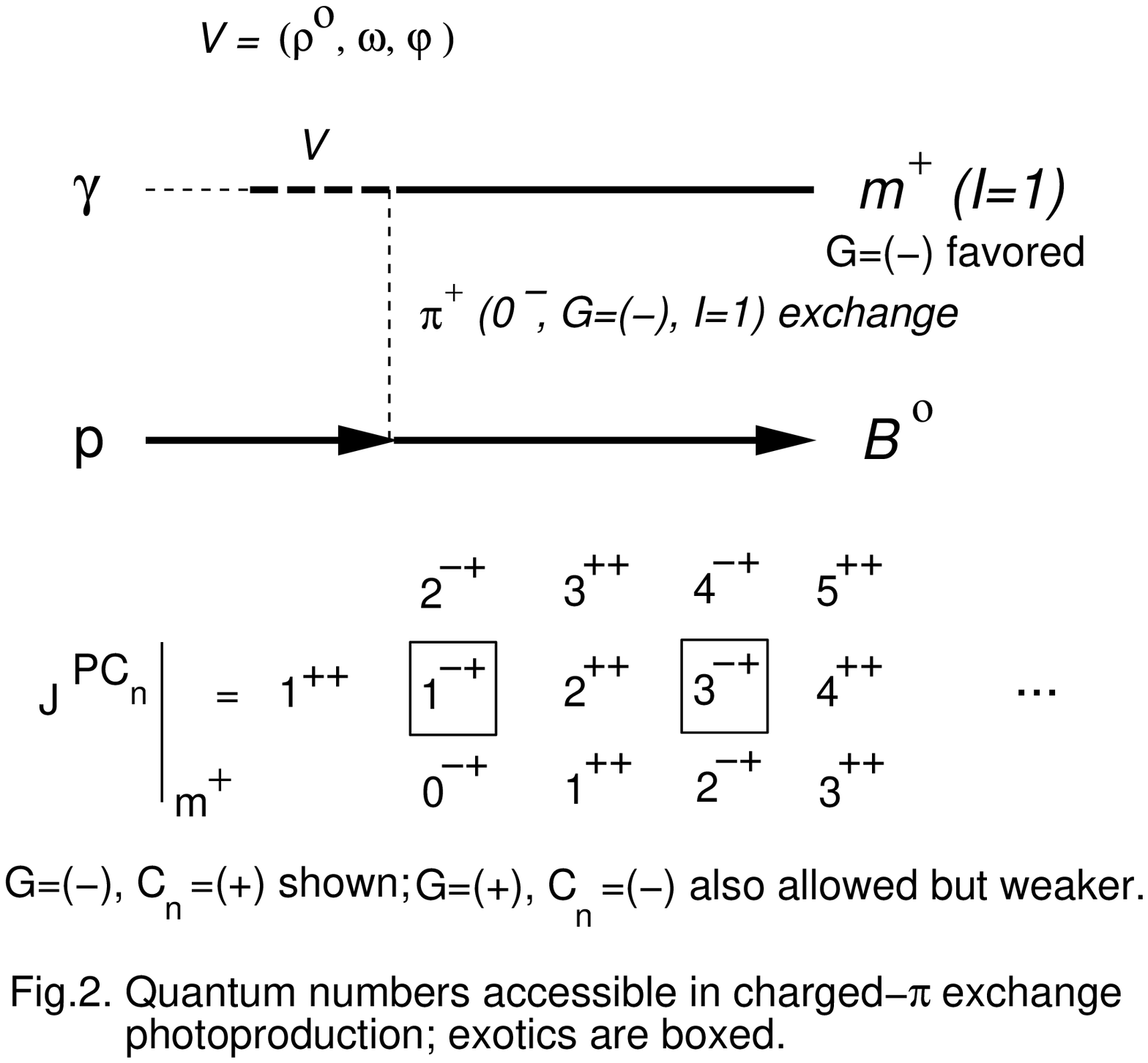,height=3in,width=3in}}
\vspace{10pt}
\end{figure}

Neutral pion exchange is unusual in that $I=0$ 
should dominate. $G=(-)$ is again
preferred, and the exotics are $even^{+-}$ and the unusual $0^{--}$.
The $2^{+-}$ is expected to be lightest, and should decay mainly to $b_1\pi$,
with $\Gamma_{tot}=0.3$ GeV. The $\rho\pi$ width of this state is 
predicted to be very small, however, so it might be difficult to photoproduce.
\begin{figure}[b!] 
\centerline{\epsfig{file=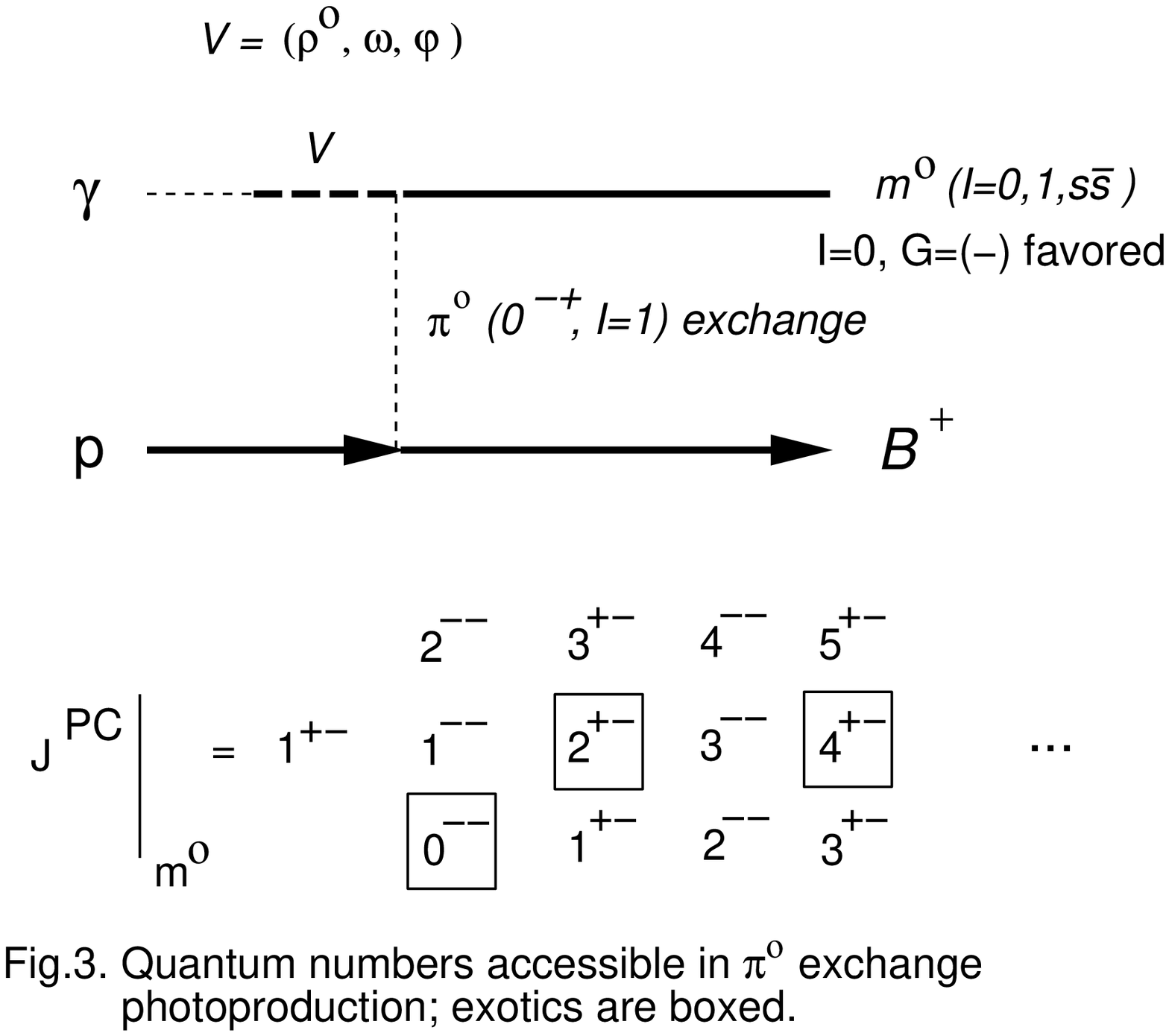,height=3in,width=3in}}
\vspace{10pt}
\end{figure}

Finally, although one prefers exotic channels because they are unambiguously
non-$q\bar q$, the flux-tube model and previous experiments
suggest several non-exotic channels for
photoproduction of hybrids.
The highest priority is the ``extra'' $\pi_J(1770)$ state reported
in photoproduction by Condo {\it et al.} \cite{Condo}, 
which may be $2^{-+}$ but is apparently
not the $\pi_2(1670)$; note that a suggestively similar
doubling of $\eta_2$ states has
been reported by Crystal Barrel~\cite{eta2}. 
The $\pi(1800)$  hybrid candidate discussed by VES \cite{VESpi1800} clearly
does not decay as the $^3$P$_0$ model predicts a 3S $q\bar q$ state should
\cite{bcps},
notably because of the S+P mode $\pi f_0(1300)$; here we may be seeing a
$0^{-+}$ hybrid or a failure of the standard $q\bar q$ decay model!
An $a_1$-type hybrid with $\Gamma_{tot}=0.5$ GeV is expected in $f_2\pi^+$
and $f_1\pi^+$.
A last case, the narrowest hybrid
predicted by the flux-tube model, is an $\omega$ state with
$\Gamma_{tot}=0.1$ GeV that should decay into $K_1K$ (both) and notably into
the final state $\omega\eta$, which is attractive experimentally.
This dramatically narrow state could serve
as a crucial test of the flux-tube picture of hybrids.

To conclude, a note of caution may be appropriate. There are no independent
tests of the flux-tube model of hybrid decays, which may be inaccurate.
Should hybrids be much broader than anticipated in some channels,
the coupling to meson continuua may give important mass shifts.
In model studies one finds that these mass
shifts are typically downwards and are numerically
comparable to the widths, so broad hybrids might
lie far from ``quenched'' theoretical predictions.
This rather drastic scenario
could explain why quenched hybrid masses in the flux-tube model and LGT
differ considerably from the masses reported for
the new E852 exotic 
candidates. In this confused situation 
the most important future experimental exercise will
be to confirm (or refute!) the existence of the light exotics
$\pi_1(1400)$ and $\pi_1(1600)$.

\subsection*{Acknowledgements}

It is a pleasure to acknowledge the kind invitation of the organisers
to present this material. I would also like to thank 
P.R.Page for technical information regarding decays of hybrids in the
flux-tube model and N.Cason, S.U.Chung, A.Dzierba, N.Isgur and E.S.Swanson
for discussions of related issues.
This work was supported in part by the USDOE
under Contract No. DE-AC05-96OR22464 managed by Lockheed Martin Energy
Research Corp.

\end{document}